\documentclass[aps,twocolumn,prb,showpacs,preprintnumbers,amsmath,amssymb]{revtex4}
\newcommand{\lsim} 
 {\ \raise.35ex\hbox{$<$}\kern-0.75em\lower.5ex\hbox{$\sim$}\ }
\newcommand{\gsim}
 {\ \raise.35ex\hbox{$>$}\kern-0.75em\lower.5ex\hbox{$\sim$}\ }

\newcommand{\ket}[1]{\left|#1\right>}
\newcommand{\mean}[1]{\left<#1\right>}
\newcommand{\means}[1]{\langle#1\rangle}

\def\journal #1#2#3#4{#1 {\bf #2}, #3 (#4)}

\def\PRA{Phys.\ Rev.\ A}
\def\PRB{Phys.\ Rev.\ B}
\def\PRL{Phys.\ Rev.\ Lett.}

\def\JPSJ{J.\ Phys.\ Soc.\ Jpn.}

\usepackage{graphicx}
\usepackage{dcolumn}
\usepackage{bm}
\usepackage{amsmath}
\usepackage{times}
\usepackage[usenames]{color}

\usepackage{ulem}

\begin{document}
\title{Finite-Temperature Phase Transition to a Quantum Spin Liquid in a Three-Dimensional Kitaev Model on a Hyperhoneycomb Lattice}
\author{J.~Nasu, T.~Kaji, K.~Matsuura$^1$, M.~Udagawa, and Y.~Motome} 
 \affiliation{Department of Applied Physics, University of Tokyo, Hongo, 7-3-1, Bunkyo, Tokyo 113-8656, Japan \\ 
$^1$Department of Advanced Materials Science, University of Tokyo, Kashiwa 277-8561, Japan}
\date{\today}
\begin{abstract}
Quantum spin liquid is an enigmatic entity that is often hard to characterize within the conventional framework of condensed matter physics. We here present theoretical and numerical evidence for the characterization of a quantum spin liquid phase extending from the exact ground state to a finite critical temperature. We investigate a three-dimensional variant of the Kitaev model on a hyperhoneycomb lattice in the limit of strong anisotropy; the model is mapped onto an effective Ising-type model, where elementary excitations consist of closed loops of flipped Ising-type variables on a diamond lattice. 
 Analyzing this effective model by Monte Carlo simulation, we find a phase transition from quantum spin liquid to paramagnet at a finite critical temperature $T_c$ accompanied with  divergent singularity of the specific heat. We also compute the magnetic properties in terms of the original quantum spins. We find that the magnetic susceptibility exhibits a broad hump above $T_c$, while it obeys the Curie law at high temperature and approaches a nonzero Van Vleck-type constant at low temperature. 
Although the susceptibility changes continuously at $T_c$, its temperature derivative shows critical divergence at $T_c$. We also clarify that the dynamical spin correlation function is momentum independent but shows quantized peaks corresponding to the discretized excitations. Although the phase transition accompanies no apparent symmetry breaking in terms of the Ising-type variables as well as the original quantum spins, we characterize it from a topological viewpoint. We find that, by defining the flux density for loops of the Ising-type variables, the transition is interpreted as the one occurring from the zero-flux quantum spin liquid to the nonzero-flux paramagnet; the latter has a Coulombic nature due to the local constraints. The role of global constraints on the Ising-type variables is examined in comparison with the results in the two-dimensional loop model. A correspondence of our model to the Ising model on a diamond lattice is also discussed. A possible relevance of our results to the recently-discovered hyperhoneycomb compound, $\beta$-Li$_2$IrO$_3$, is mentioned.
\end{abstract}



\maketitle



%
%

%




\section{Introduction}

Quantum spin liquid (QSL) is one of the central issues in condensed matter physics.~\cite{Balents2010} This is a new state of matter in insulating magnets, which shows no apparent symmetry breaking. Experimental candidates for QSL were recently discovered in several quasi-two- and three-dimensional (3D) compounds, which have been stimulating the study of QSL.~\cite{Shimizu03,Nakatsuji05,Helton07,Okamoto07,Yamashita10} In these compounds, the identification of QSL often relies on the lack of singularity in thermodynamic quantities, especially, the absence of magnetic long-range ordering down to the lowest temperature ($T$). On the other hand, theoretically, in addition to a number of one-dimensional systems, several solvable models have been proposed and served as prototypes for QSL in higher dimensions.~\cite{Rokhsar1988,Moesner2001,Kitaev06} In addition, intensive numerical researches have also been done to explore QSL.~\cite{Morita2002,Yan2011,Jiang2012}

A fundamental question in the research of QSL is how to characterize QSL and distinguish it from a simple paramagnet. In general, liquid and gas have the same symmetry and they are not necessarily distinguished by a phase transition. In classical fluids, there is a first-order transition between liquid and gas, but the discontinuous phase boundary is terminated at a critical endpoint; beyond that, liquid and gas are adiabatically connected with each other by a crossover. Then, what is the case of a quantum fluid, QSL? Is the high-$T$ paramagnetic phase adiabatically connected to the low-$T$ QSL phase? Since there is no apparent symmetry breaking between these two phases as in the case of the classical liquid and gas, it is naively expected that QSL and paramagnet are adiabatically connected. On the other hand, some QSL phases are characterized by a topological order related with the ground state degeneracy or finite topological entropy, while the paramagnetic phase is not.~\cite{Wen_textbook} This implies the existence of some singularity related to the topological order, but it is not clear whether the singularity accompanies a conventional phase transition. The question is not only purely theoretical but also relevant to the interpretation of experimental results in the QSL candidates.

The issue of the connection between QSL and paramagnetic state has been recently studied in the Kitaev model and its generalizations. The advantage to adopt the models is the availability of exact solutions in the excited states as well as in the ground state. The original Kitaev model defined on a two-dimensional (2D) honeycomb lattice is exactly solvable by utilizing a local Ising-type conserved quantity on each hexagon of the honeycomb lattice,~\cite{Kitaev06} and shows a QSL ground state with short-range spin correlation.~\cite{Baskaran2007} Meanwhile, when one type of three inequivalent bonds is much stronger than the other two, the model is mapped onto the toric code model: one of the prototypical models for QSL, where the existence of topological order was exactly proved.~\cite{Hamma2005,Castelnovo2007} In the toric code limit, the effect of temperature was studied in both 2D and 3D,~\cite{Nussinov2008,Castelnovo2007,Castelnovo2008,Iblisdir2009,Iblisdir2010} and in some cases, the topological order was shown to persist up to the critical temperature. It is highly desired to explore such expanding forefront for deeper understanding of QSL physics.

In this paper, we present convincing evidence of a finite-$T$ phase transition between QSL and paramagnet in a 3D variant of the Kitaev model. This 3D Kitaev model, which was originally introduced in Ref.~\onlinecite{Mandal2009}, is defined on a 3D hyperhoneycomb lattice, and inherits the solvability of the 2D honeycomb counterpart in the ground state. The ground-state phase diagram has the same structure as that in 2D; the model provides an example of 3D QSL. However, the limit of one stronger bond than the other two leads to a distinct model from the 2D toric code model; the excitations are allowed only in the form of peculiar closed loops because of the local constraints on the Ising-type variables, which makes the thermodynamics nontrivial. Performing the numerically exact analysis by Monte Carlo (MC) simulation, we show that, although the model is apparently noninteracting, it exhibits a phase transition at a finite critical temperature $T_c$ between QSL and paramagnetic phases because of the local constraints. We find that $T$ dependence of the internal energy largely deviates from the noninteracting one and that the specific heat exhibits a divergent singularity at $T_c$. A large amount of entropy is released in the high-$T$ paramagnetic phase. Moreover, we successfully calculate the magnetic susceptibility in terms of the original quantum spins. We show that the susceptibility obeys the Curie law at high $T$, while it exhibits a broad hump above $T_c$. Although the susceptibility decreases continuously at $T_c$, its $T$ derivative shows divergent singularity. A nonzero Van Vleck-type contribution appears below $T_c$. 
We also calculate the dynamical spin correlation function at finite $T$. It is momentum independent due to the extremely short-range spin correlation, but shows quantized peaks as a function of excitation energy. The peak at the highest energy rapidly increases near $T_c$ as $T$ decreases, while other peaks are suppressed.

The phase transition is hard to characterize in terms of the local variables, as it apparently accompanies no symmetry breaking. We characterize it by the emergence of extended loops of flipped Ising-type variables. We define the flux for each loop and successfully identify the transition from the paramagnetic side in terms of the flux density. 
We find that the transition is characterized by the emergence of flux in the high-$T$ paramagnetic state which has a Coulombic nature due to the local constraints. 
We also show that the finite-size scaling of the flux density suggests the critical exponents consistent with the 3D Ising universality class.

Meanwhile, in addition to the local constraints, there are global constraints on the Ising-type variables.
We show that the global constraints are not relevant to the thermodynamics and the model omitting them also shows the same phase transition. 
The transition belongs to the 3D Ising universality class because the effective model without the global constraints corresponds to an Ising model on a diamond lattice. Although this phase transition is continuous, we find that the parameter which describes the topology of loops is expected to show a discontinuous jump at $T_c$ in the thermodynamic limit. Moreover, we analytically show that this parameter is given by a step function in a 2D loop model.
We also mention a possible relation to a recently found compound, $\beta$-Li$_2$IrO$_3$.

The paper is structured as follows. In Sec.~\ref{sec:model}, we introduce a 3D variant of the Kitaev model defined on a hyperhoneycomb lattice and its effective model in the limit of strong anisotropy. In Sec.~\ref{sec:method}, we present a numerical method to solve the effective model, MC simulation by the Wang-Landau algorithm.~\cite{WL} In Sec.~\ref{sec:results}, we show the results of our numerical analysis. $T$ dependence of the energy, specific heat, and entropy is shown in Sec.~\ref{sec:energy}. We identify a finite-$T$ phase transition accompanied by a divergent singularity of the specific heat. In Sec.~\ref{sec:magn-susc}, the magnetic susceptibility is calculated and its characteristic $T$ dependence is discussed in comparison with the 2D result. We also compute the dynamical spin correlation function and discuss its relation to the static susceptibility in Sec.~\ref{sec:dynam-corr-funct}. An attempt to characterize the phase transition from a topological viewpoint is discussed by examining the flux density defined by the loop excitations in Sec.~\ref{sec:loop-configuration}. In Sec.~\ref{sec:corr-ising-model}, we examine the role of global constraints in comparison with a 2D loop model, and also discuss a correspondence of our effective model to an Ising model on a diamond lattice.
Finally, Sec.~\ref{sec:concluding-remarks} is devoted to concluding remarks. Details of the calculation for the Van Vleck-type contribution in the magnetic susceptibility are given in Appendix~\ref{sec:derivation-van-vleck}. The analysis of the 2D loop model is presented in Appendix~\ref{sec:2d-loop-model}.


\section{Model}\label{sec:model}

\begin{figure}[t]
\begin{center}
\includegraphics[width=\columnwidth,clip]{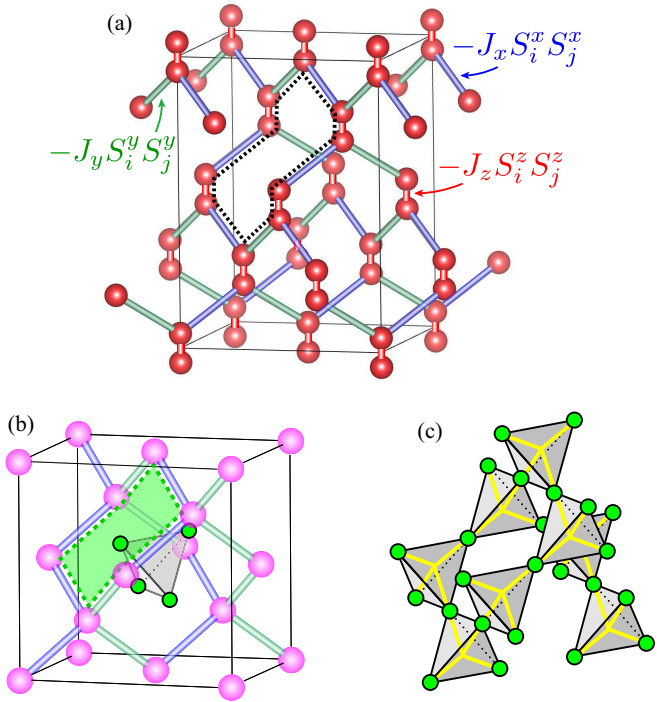}
\caption{(Color online) 
(a) Schematic picture of the hyperhoneycomb lattice structure and interactions in the 3D Kitaev model in Eq.~(\ref{eq:1}). The dotted line represents a ten-site loop on which the conserved quantity $K_p$ in Eq.~(\ref{eq:8}) is defined. (b) Schematic picture of the diamond lattice which is formed by contracting the dimers of $z$-bonds on the hyperhoneycomb lattice. The dotted hexagon represents a six-site loop corresponding to the ten-site loop in (a). The centers of four six-site loops sharing their edges are represented by small circles, which constitute a tetrahedral primitive cell of the pyrochlore lattice (see the main text and Fig.~\ref{fig:pyrochlore}). (c) Schematic picture of the pyrochlore lattice on which the effective Hamiltonian in Eq.~(\ref{eq:2}) is defined. The corresponding loop model is defined on the diamond lattice composed of the centers of tetrahedra connected by thick yellow lines.
}
\label{fig:kitaev}
\end{center}
\end{figure}

\begin{figure}[t]
\begin{center}
\includegraphics[width=\columnwidth,clip]{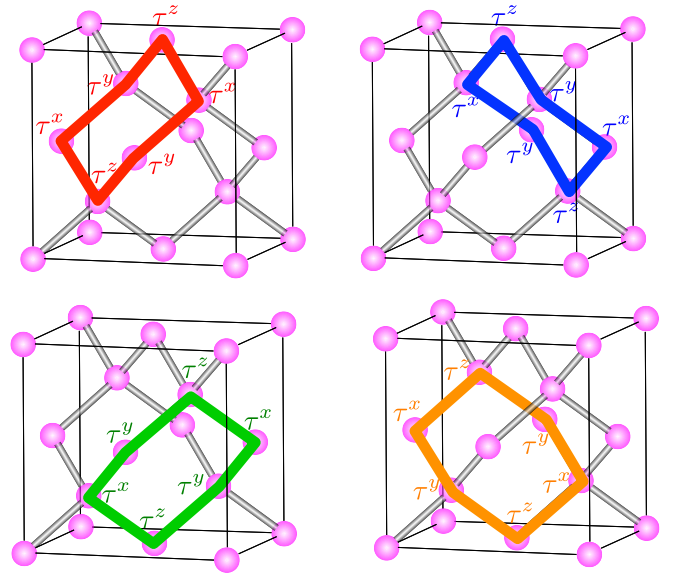}
\caption{(Color online) 
Four kinds of inequivalent six-site loops on which $B_p$ are defined. Each $B_p$ is given by the product of six $\tau_m^l$ which are shown in the figure. See also Eq.~(\ref{eq:B_p}). 
}
\label{fig:diamond}
\end{center}
\end{figure}

\begin{figure}[t]
\begin{center}
\includegraphics[width=\columnwidth,clip]{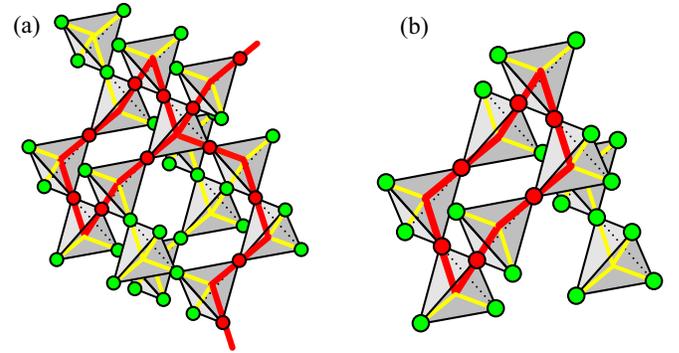}
\caption{(Color online) 
(a) A loop configuration on the diamond lattice shown in Fig.~\ref{fig:kitaev}(c). Light (dark) circles represent sites with $B_p=+1$ $(-1)$. (b) The shortest loop excitation consisting of six sites.
}
\label{fig:pyrochlore}
\end{center}
\end{figure}

We consider the Kitaev model on the 3D lattice depicted in Fig.~\ref{fig:kitaev}(a), which is called a hyperhoneycomb lattice. This lattice is composed of three kinds of nearest-neighbor (NN) bonds termed $x$-, $y$-, and $z$-bonds. The Kitaev Hamiltonian is defined on this lattice in the form of
\begin{align}
 {\cal H}=-J_x\sum_{\means{ij}_x}S_i^x S_j^x-J_y\sum_{\means{ij}_y}S_i^y S_j^y-J_z\sum_{\means{ij}_z}S_i^z S_j^z,
 \label{eq:1}
\end{align}
where $S_i^l$ represents the $l$ component of spin-$1/2$ operator, and $\means{ij}_l$ denotes a NN $l$-bond ($l=x,y,z$). Here, $J_x$, $J_y$, and $J_z$ are set to be positive. The $z$-bond is crystallographically inequivalent to the $x$- and $y$-bonds on the hyperhoneycomb lattice, implying $J_z\not=J_x=J_y$.

The model in Eq.~(\ref{eq:1}) is topologically equivalent to that introduced in Ref.~\onlinecite{Mandal2009}, whose fundamental properties at $T=0$ have been discussed. In particular, the ground state phase diagram is exactly obtained in
the same structure as that for the Kitaev model on a honeycomb lattice.~\cite{Kitaev06} This is owing to the local conserved quantities commonly existing in the two models. In the hyperhoneycomb case, the conserved quantity is defined by
\begin{align}
 K_p=\prod_{n=1}^{10}\sigma_{i_n}^{l_{i_n}}\label{eq:8}
\end{align}
 on each ten-site loop $p$, as exemplified in Fig.~\ref{fig:kitaev}(a). Here, $\sigma_i^l=2S_i^l$ is the $l$ component of the Pauli matrices, $i_n$ represents a site belonging to a ten-site loop $p$, and $l_i$ is defined as the bond not on the loop $p$ among the three NN bonds connected at the site $i$.

In the present study, we focus on the limit of $J_z\gg J_x$, $J_y$, and set $J_x=J_y=J$ for simplicity. This anisotropy does not break the symmetry of the hyperhoneycomb lattice. In this limit, it was shown that the low-energy physics of the model in Eq.~(\ref{eq:1}) can be described by an effective Ising-type model,~\cite{Mandal2009} as briefly reviewed below. 
For $J_z>0$ and $J=0$, the lattice is decomposed into the dimers of $z$-bonds, and the spins form independent doublets on each dimer: $\ket{\Uparrow}\equiv\ket{\uparrow\uparrow}$ and $\ket{\Downarrow}\equiv\ket{\downarrow\downarrow}$, as far as the temperature region $T\ll J_z$ is concerned. For each dimer $m$, it is convenient to introduce the pseudo-spin (PS) operators $\bm{\tau}_m$, which are the Pauli matrix operators satisfying $\tau^z_m \ket{\Uparrow} = \ket{\Uparrow}$, $\tau^z_m\ket{\Downarrow} = -\ket{\Downarrow}$, $\tau^x_m\ket{\Uparrow} = \ket{\Downarrow}$, and $\tau^x_m\ket{\Downarrow} = \ket{\Uparrow}$. 
The product states of $\ket{\Uparrow}$ or $\ket{\Downarrow}$ give macroscopically degenerate ground states.
An infinitesimally small $J$ connects these dimers, and lifts this degeneracy. 
The perturbation in terms of $J/J_z$ leads to multiple spin interactions induced by a ring-exchange-type processes.
Since the dimers compose a diamond lattice shown in Fig.~\ref{fig:kitaev}(b), 
the leading-order contribution comes from the shortest
six-site loops $p$ on the diamond lattice, which correspond to ten-site loops $p$ on the original hyperhoneycomb lattice (see Fig.~\ref{fig:kitaev}). 
This process leads to an effective Hamiltonian;
\begin{align}
{\cal H}_{\rm eff}=-J_{\rm eff}\sum_p B_p,
\label{eq:2}
\end{align}
where the effective coupling constant $J_{\rm eff}$ is given by $7J^6/(1024J_z^5)$.~\cite{Mandal2009} Here, the summation is taken over the loops $p$, whose centers comprise a pyrochlore lattice [see Figs.~\ref{fig:kitaev}(b) and \ref{fig:kitaev}(c)]. The multiple spin interaction is compactified into the form of conserved operator, $B_p={\cal P}K_p{\cal P}$, where ${\cal P}$ is the projection to the states spanned by $\ket{\Uparrow}$ and $\ket{\Downarrow}$;
\begin{align}
B_p = \prod_{m=1}^{6} \tau_m^l,
\label{eq:B_p}
\end{align}
where $m$ runs on the six-site loop $p$ on the diamond lattice. 
There are four different $B_p$ depending on the types of original ten-site loops, as shown in Fig.~\ref{fig:diamond}; for each $B_p$, $\tau_m^l$ is given by the rule shown in Fig.~\ref{fig:diamond}. As the operators $B_p$ commute with each other and each $B_p$ takes the value $\pm 1$ ($B_p^2=1$), the eigenstates of the Hamiltonian in Eq.~(\ref{eq:2}) are labeled by a set of values of the Ising-type variables $B_p$. Hence, the effective model in Eq.~(\ref{eq:2}) is the Ising-type model defined on the pyrochlore lattice.~\footnote{
Each eigenstate labeled by the values of $B_p$ is eight-fold degenerate, but the degeneracy trivially gives a constant shift for the free energy.~\cite{Mandal2011} Hence, we ignore this degeneracy in the following analysis.}

The ground state of the Hamiltonian in Eq.~(\ref{eq:2}) is trivially given by the state with $B_p=+1$ for all $p$. In excited states, $B_p$ takes $-1$ for several $p$. In particular, the lowest excitation appears to be given by the states with a single flip, $B_p=+1 \to -1$. The variables $B_p$, however, are not independent of each other, and such a single flip is prohibited as explained below. As the original Pauli operators $\sigma_i^{l_i}$ on adjacent four ten-site loops are multiplied to give $1$ as an operator identity, the four $B_p$ shown in Fig.~\ref{fig:diamond} are multiplied to be $1$. In terms of the pyrochlore lattice in Fig.~\ref{fig:kitaev}(c), such constraint is summarized as $\prod_{p\in T_i}B_p=1$ for all the tetrahedra $T_i$. This immediately means that only the excited states where the sites with $B_p=-1$ form closed loops are allowed [see Fig.~\ref{fig:pyrochlore}(a)].

Consequently, the effective Hamiltonian in Eq.~(\ref{eq:2}) is considered as a loop model on the diamond lattice obtained by connecting the centers of neighboring tetrahedra in the pyrochlore lattice~\cite{Mandal2011} [see Fig.~\ref{fig:kitaev}(c)].
Note that $B_p$ variables are located on the bonds of the diamond lattice and that this diamond lattice is different from that for the PS operators $\bm{\tau}_m$ in Fig.~\ref{fig:kitaev}(b). In this representation, the energy is proportional to the sum of the loop lengths. The loops with $B_p=-1$ can intersect with each other, as shown in Fig.~\ref{fig:pyrochlore}(a). The shortest loop is a six-site loop $C_s$ shown in Fig.~\ref{fig:pyrochlore}(b). All of the loop configurations can be constructed by the ``product'' $C_{s_1}\otimes C_{s_2}\otimes C_{s_3}\otimes \cdots$, where $C_{s_i}\otimes C_{s_j}\equiv C_{s_i}\cup C_{s_j}-C_{s_i}\cap C_{s_j}$. Moreover, in addition to the local constraints in all the tetrahedra, there are two kinds of global constraints:~\cite{Mandal2011} (i) the number of the points at which loops intersect with each $xy,yz,zx$ plane is even, and (ii) the loop configurations are constructed by the even-number ``product'' of the six-site loops.

In the following sections, we investigate thermodynamic properties of the effective model given by Eq.~(\ref{eq:2}). Let us make two remarks on the model. 
One is about fermionic excitations. In the original Hamiltonian given in Eq.~(\ref{eq:1}), there is fermionic degree of freedom in each state labeled by the local conserved quantities $K_p$.\cite{Mandal2009} In the effective Ising-type model in Eq.~(\ref{eq:2}), the fermionic excitations correspond to the excitations to the states projected out in the dimer representation. Thus, they can be neglected in the thermodynamics as well as the ground-state properties for the model in the limit of $J\ll J_z$. 
The other remark is on the higher-order perturbations. As mentioned above, the effective model in Eq.~(\ref{eq:2}) includes only the leading-order controbution from the perturbation in terms of $J/J_z$. The higher-order terms give rise to complicated interactions, which induce interactions between the loops of $B_p$ variables. We, however, believe that the loop structure is retained even if the higher-order perturbations are taken into account, because the local constraints for the conserved quantity $B_p$ originate from the algebraic property of $\tau_m^l$. We will further discuss this point in the end of Sec.~\ref{sec:loop-configuration}.


\section{Method}\label{sec:method}

Although the model in Eq.~(\ref{eq:2}) apparently describes free Ising moments in applied magnetic field $J_{\rm eff}$, the constraints give rise to intersite correlations between $B_p$. This makes the finite-$T$ behavior nontrivial in the current 3D model. Note that the 2D toric code model is free from such constraints, and hence, shows no phase transition at finite $T$. In order to examine the thermodynamic properties, particularly, the existence of finite-$T$ phase transition, we adopt an unbiased numerical method, MC simulation by the Wang-Landau algorithm.~\cite{WL} The simulations were performed on the pyrochlore lattice with $N=4\times L^3$ sites under periodic boundary conditions. The system size is up to $L=30$. After obtaining the density of states, $10^8$ MC steps are spent for standard measurement. By performing 20-40 times iterations of this measurement independently, we obtained physical quantities with their statistical errors. We performed MC sampling by flipping a pair of six-site loops at once in each update, which automatically satisfies both global constraints (i) and (ii) as well as the local constraints. For comparison, we will show the results when we neglect the global constraints in Sec.~\ref{sec:corr-ising-model}. Hereafter, we take $J_{\rm eff}=1$ as an energy scale and the Boltzmann constant $k_{\rm B}=1$.
We choose the lattice constant of the pyrochlore lattice to be unity.


\section{Results}\label{sec:results}

\subsection{Energy and Specific Heat}\label{sec:energy}

\begin{figure}[t]
\begin{center}
\includegraphics[width=\columnwidth,clip]{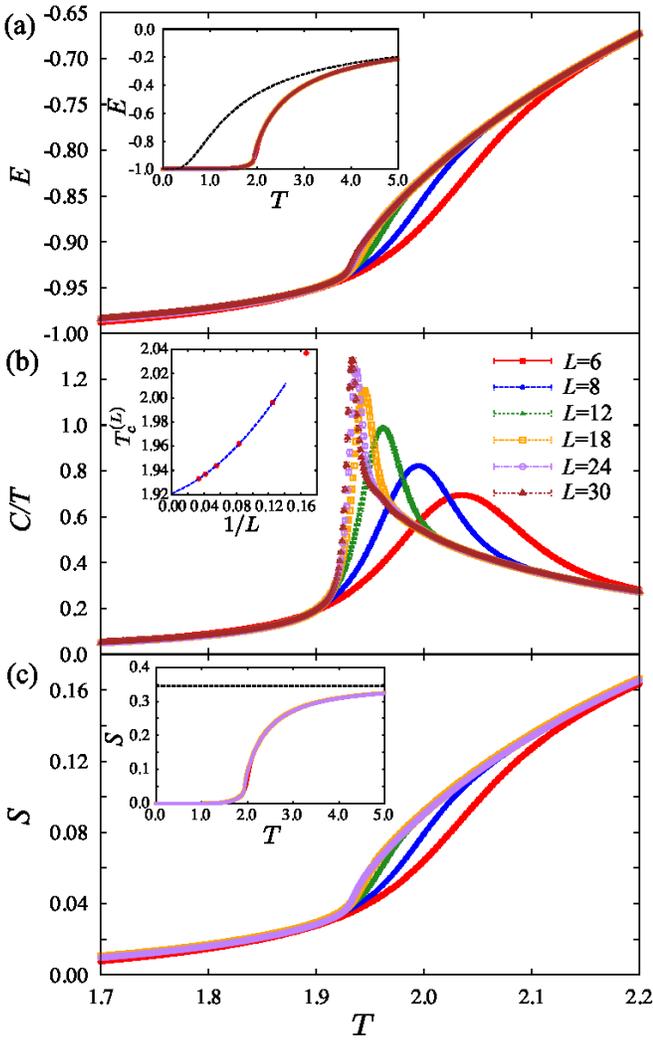}
\caption{(Color online) 
MC results for (a) $E$, (b) $C/T$, and (c) $S$ as functions of $T$. Insets in (a) and (c) are the plots in a wide $T$ range. The dotted curve in the inset of (a) shows $-\tanh(T^{-1})$ for comparison. The inset in (b) shows the peak temperature of $C$ as a function of $1/L$. The dotted curve represents the quadratic fit for 5 largest $L$. The dotted horizontal line in the inset of (c) shows the value of $\frac12 \ln 2$.
}
\label{fig:mc}
\end{center}
\end{figure}

The numerical results for the effective model in Eq.~(\ref{eq:2}) are presented in Fig.~\ref{fig:mc}. Figure~\ref{fig:mc}(a) shows the internal energy per site defined by
\begin{align}
 E=\frac{1}{N}\means{{\cal H}_{\rm eff}},
\end{align}
where $\means{\cdots}$ denotes the thermal average.
As shown in the inset of Fig.~\ref{fig:mc}(a), $T$ dependence of $E$ largely deviates from $-\tanh(T^{-1})$ that is expected for a free Ising moment in a magnetic field $J_{\rm eff}=1$; $E$ does not change so much up to $T =T_c \simeq 1.92$, while it rapidly increases for $T>T_c$. The deviation is due to the constraints on the Ising-type variables $B_p$, and the kink behavior around $T_c$ suggests a phase transition. Indeed, $E$ shows a significant system-size dependence, as shown in Fig.~\ref{fig:mc}(a). Accordingly, the specific heat, calculated by 
\begin{align}
C = \frac{1}{NT^2}(\langle {\cal H}_{\rm eff}^2 \rangle - \langle {\cal H}_{\rm eff} \rangle^2),
\end{align}
exhibits a sharp peak around $T_c$ which grows with increasing the system size; $C$ divided by $T$ is plotted in Fig.~\ref{fig:mc}(b). The size dependence of the peak temperature of $C$, $T_c^{(L)}$, is presented in the inset of Fig.~\ref{fig:mc}(b). The extrapolation of $T_c^{(L)}$ to $L\to \infty$ gives an estimate of the critical temperature as $T_c=1.921(1)$. The results indicate that there is a phase transition at $T=T_c$. The transition is not discontinuous; we see no sign of a double peak in the energy histograms (not shown).

Figure~\ref{fig:mc}(c) shows the result for the entropy per site, which is calculated by the numerical integration, 
\begin{align}
S= \int_0^T \frac{C}{T'} dT'.
\end{align}
$S$ appears to saturate at a much smaller value than $\ln 2$ at $T\to \infty$, as shown in the inset. 
The saturation value is close to $\frac{1}{2}\ln 2\sim 0.347$.
This is due to the local constraint which allows $B_p$ to take only a half of $2^4$ configurations on every tetrahedron. 
Note that the value of the entropy corresponds to the degree of freedom in the space of the PS $\bm{\tau}_m$ on the diamond lattice; the number of sites on the diamond lattice in Fig.~\ref{fig:kitaev}(b) is a half of that on the pyrochlore lattice in Fig.~\ref{fig:kitaev}(c) as well as the original hyperhoneycomb lattice in Fig.~\ref{fig:kitaev}(a). Interestingly, most of the entropy is released not at the transition but in the high-$T$ phase while decreasing $T$. This suggests a strong short-range correlation between the variables $B_p$ even well above $T_c$. 
This corresponds to the development of short-range correlations in the variables $B_p$ even well above $T_c$.
Such correlations are indicated in the dynamical spin correlation function calculated in Sec.~\ref{sec:dynam-corr-funct}. 
They are also seen in terms of loops; the high-$T$ phase has a Coulombic nature due to the local constraints, as discussed in Sec.~\ref{sec:loop-configuration}.

\subsection{Magnetic susceptibility}\label{sec:magn-susc}

\begin{figure}[t]
\begin{center}
\includegraphics[width=\columnwidth,clip]{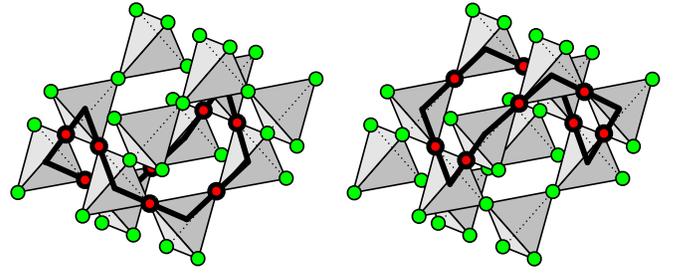}
\caption{(Color online) 
Two kinds of inequivalent eight-site loops of $B_p$ whose $p \in {\cal A}_m$ (thick lines). These loops contribute the magnetic susceptibility in Eq.~(\ref{eq:chi^zz_Q}). See the text for details.
}
\label{fig:eight_loop}
\end{center}
\end{figure}

\begin{figure}[t]
\begin{center}
\includegraphics[width=0.85\columnwidth,clip]{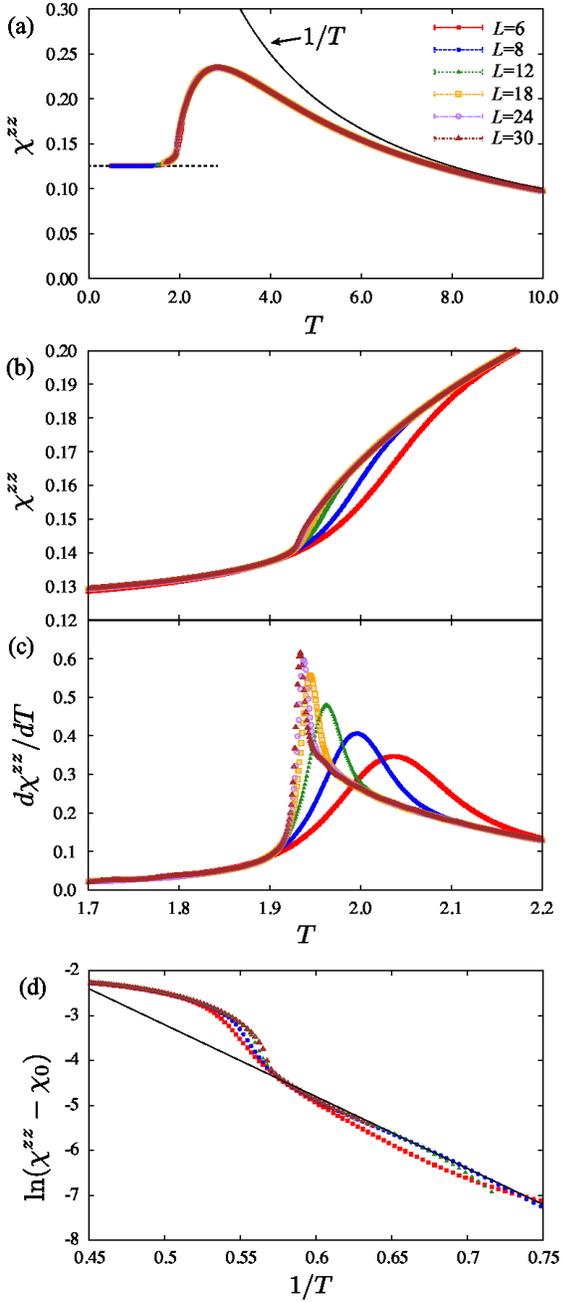}
\caption{(Color online) 
MC results for (a) the magnetic susceptibility $\chi^{zz}$ and (b) its narrow $T$ range plot near $T_c$.
In (a), the curve shows the Curie behavior $1/T$, and the horizontal dotted line represents the Van Vleck-type component $\chi_0=1/8$. 
(c) $T$ derivative $d\chi^{zz}/dT$. 
(d) Arrhenius plot for $\chi^{zz}-\chi_0$. The line shows a linear function whose slope is $-16$.
}
\label{fig:suscep}
\end{center}
\end{figure}

In this section, we calculate the magnetic susceptibility in terms of the original spins $\bm{S}_i$ in Eq.~(\ref{eq:1}) in magnetic field applied parallel to the $z$ direction. According to the Kubo formula, the susceptibility is calculated as
\begin{eqnarray}
\chi_s^{zz}=\frac{1}{N}\sum_{ij}
\int_0^\beta d\lambda \means{e^{\lambda{\cal H}}
S_i^z e^{-\lambda{\cal H}}
S_j^z
}.
\end{eqnarray}
In the limit of $J_z \gg J$, $\chi^{zz}$ can be expressed by PS operators $\bm{\tau}_m$ as
\begin{align}
 \chi^{zz}=2\chi_s^{zz}=\frac{1}{N_d}\sum_{mn}
\int_0^\beta d\lambda \means{e^{\lambda{\cal H}_{\rm eff}}
\tau_m^z e^{-\lambda{\cal H}_{\rm eff}}
\tau_n^z
}.\label{eq:5}
\end{align}
Here, we introduce the factor 2 to renormalize $\chi^{zz}$ by $N_d=N/2$ which is the site number of the diamond lattice composed of the dimers of $z$-bonds [see Fig.~\ref{fig:kitaev}(b)].

In order to calculate the susceptibility in Eq.~(\ref{eq:5}), we introduce $\tilde{\cal H}_{\rm eff}^m$ so as to satisfy the relation $\tau_m^z{\cal H}_{\rm eff}=\tilde{\cal H}_{\rm eff}^m\tau_m^z$. For a given dimer site $m$, the operator $\tau_m^z$ commutes with all $B_p$ that do not involve the site $m$ in the loop $p$. On the other hand, for $B_p$ involving the site $m$, the commutation relation between $B_p$ and $\tau_m^z$ depends on the location of $m$ on the loop $p$ as follows. All $B_p$ include two $\tau_m^z$, as shown in Fig.~\ref{fig:diamond}. We term such two sites $m_1^p$ and  $m_2^p$. If $m$ is equal to either $m_1^p$ or $m_2^p$, $B_p$ commutes with $\tau_m^z$. Otherwise, $B_p$ anticommutes with $\tau_m^z$. The number of $B_p$ that anticommute with the given operator $\tau_m^z$ is eight; the eight $B_p$ form a loop. There are two kinds of such eight-site loops, as shown in Fig.~\ref{fig:eight_loop}. We define the set of $p$ for such $B_p$ as ${\cal A}_m$. From the definition of ${\cal A}_m$, $\tilde{\cal H}_{\rm eff}^m$ is written by
\begin{align}
 {\cal H}_{\rm eff}-\tilde{\cal H}^m_{\rm eff}=-2 \sum_{p\in{\cal A}_m}B_p=-2 {\cal Q}_m,\label{eq:4}
\end{align}
where ${\cal Q}_m$ is the sum of $B_p$ belonging to ${\cal A}_m$. Note that ${\cal Q}_m$ takes the values $-8, -6, -4, \cdots, 8$. Then, the magnetic susceptibility in Eq.~(\ref{eq:5}) is rewritten as
\begin{align}
\chi^{zz}=\frac{1}{N_d}\sum_{mn}
\int_0^\beta d\lambda \means{e^{-2 \lambda{\cal Q}_m}
\tau_m^z 
\tau_n^z
}=\frac{1}{N_d}\sum_{m}\means{f({\cal Q}_m)},
\label{eq:chi^zz_Q}
\end{align}
where
\begin{align}
f({\cal Q}_m)&=
\int_0^\beta d\lambda e^{-2 \lambda{\cal Q}_m}
\label{eq:fQ}\\
&=\begin{cases}
\frac{1}{2 {\cal Q}_m}(1-e^{-2\beta  {\cal Q}_m})
& ({\cal Q}_m\neq 0)\\
\beta &  ({\cal Q}_m = 0)
\end{cases}.
\end{align}
In Eq.~(\ref{eq:chi^zz_Q}), we used the fact that there is no spatial correlation between $\tau^z$, i.e., $\means{\tau_m^z \tau_n^z} = \delta_{mn}$, where $\delta_{mn}$ is the Kronecker delta. This is because, in the original Kitaev model given in Eq.~(\ref{eq:1}), the equal-time spin correlation $\means{S_i^l S_j^l}$ is nonzero only if the sites $i$ and $j$ are on the same NN $l$-bond due to the existence of the local conserved 
quantity $K_p$.~\cite{Baskaran2007}

The numerical result for the magnetic susceptibility is presented in Fig.~\ref{fig:suscep}(a). In the high-$T$ region, the susceptibility obeys the Curie law rather than the Curie-Weiss law, i.e., Curie-Weiss temperature is exactly zero. 
The vanishing Curie-Weiss temperature is due to the absence of correlation between $\tau^z$ mentioned above. It is worthy noting that, as the effective model in Eq.~(\ref{eq:2}) is justified for $T\ll J_z$, the Curie law behavior is considered to be seen in the temperature range $J\ll T\ll J_z$; in the original Kitaev model in Eq.~(\ref{eq:1}), $\chi^{zz}$ should obey the Curie-Weiss law with a positive Weiss temperature proportional to $J_z$ for $T\gg J_z$. 
The anticipated crossover at $T\sim J_z$ is due to the fermion excitations which are prohibited in the effective model in Eq.~(\ref{eq:2}).

While decreasing $T$, $\chi^{zz}$ deviates from the Curie law, and shows a broad hump at $T\simeq 3$. Below the hump, it rapidly decreases toward $T_c$ and does not strongly depend on $T$ below $T_c$. In the low-$T$ limit, $\chi^{zz}$ converges to a nonzero 
constant. Since the ground state is not degenerate in terms of $B_p$, the low-$T$ behavior of the susceptibility is interpreted as the Van Vleck-type paramagnetism. As explained in Appendix~\ref{sec:derivation-van-vleck}, the asymptotic value for $T\to 0$ is analytically obtained to be $1/8$. Our result in Fig.~\ref{fig:suscep}(a) is consistent with this estimate.

As shown in Fig.~\ref{fig:suscep}(b), the susceptibility changes continuously in the vicinity of $T_c$, while showing substantial system-size dependence, similar to the internal energy and entropy shown in Figs.~\ref{fig:mc}(a) and \ref{fig:mc}(c), respectively. Figure~\ref{fig:suscep}(c) shows $T$ derivative of the susceptibility near $T_c$. A sharp peak which grows with increasing the system size is observed around $T_c$. This implies divergent behavior of $d\chi^{zz}/dT$ at $T_c$.

We also analyze the behavior of $\chi^{zz}$ below $T_c$ by the Arrhenius plot for $\chi^{zz}-\chi_0$, where $\chi_0=1/8$ is the Van Vleck-type component. As shown in Fig.~\ref{fig:suscep}(d), in the low-$T$ region, the slope of the Arrhenius plot is about $-16$. This result indicates that susceptibility except for the Van Vleck-type component is of activation type with an energy gap $\simeq 16$. This value corresponds to the energy for the lowest excitation in this gapped phase. In the current case, the lowest excitation is not a shortest six-site loop but an eight-site loop in terms of $B_p$: the lowest excitation energy is $8\times 2=16$. This is due to the global constraint (ii) given in Sec.~\ref{sec:model}; six-site loops should be flipped in pair, as implemented in our MC calculations (see also Sec.~\ref{sec:method}).

\begin{figure}[t]
\begin{center}
\includegraphics[width=\columnwidth,clip]{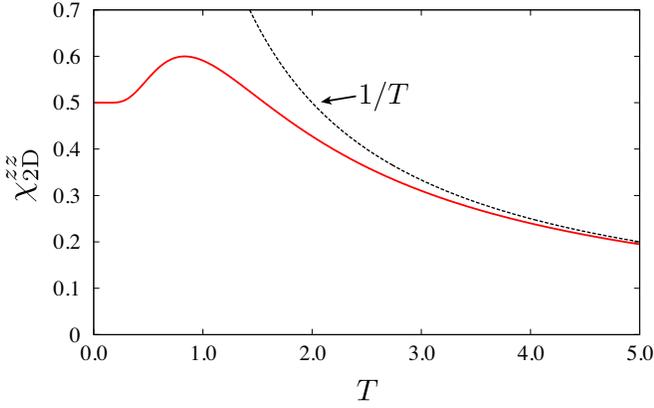}
\caption{(Color online) 
$T$ dependence of the magnetic susceptibility in the toric code limit of the Kitaev model on a honeycomb lattice. The curve shows the Curie behavior $1/T$.
}
\label{fig:suscep2D}
\end{center}
\end{figure}

Let us compare the result with that in the 2D system, for which the susceptibility can be calculated analytically. In the toric code limit $J_z \gg J_x, J_y$ for the Kitaev model on the honeycomb lattice, the effective Hamiltonian is given as
\begin{align}
 {\cal H}_{\rm 2D}=-J_{\rm 2D}\sum_p \tilde{B}_p,\label{eq:10}
\end{align}
where $\tilde{B}_p=\pm 1$ is the local conserved quantity similar to $B_p$ in the 3D case.~\cite{Kitaev06} In the 2D case, since there is no local constraint for $\tilde{B}_p$ unlike the 3D case, the correlation between $\tilde{B}_p$ does not appear. Then, the thermal average of $\tilde{B}_p$ is simply given by the noninteracting form, $\means{\tilde{B}_p}=\tanh\beta J_{\rm 2D}$. By using this result, the magnetic susceptibility $\chi_{\rm 2D}^{zz}$ defined in the same way as
Eq.~(\ref{eq:5}) is calculated in the form
\begin{align}
 \chi_{\rm 2D}^{zz}=\frac{1}{2 J_{\rm 2D}}\left(\frac{\beta J_{\rm 2D}}{\cosh^2 \beta J_{\rm 2D}}+\tanh \beta J_{\rm 2D}\right).
\end{align}
The functional form is plotted in Fig.~\ref{fig:suscep2D} by setting $J_{\rm 2D}=1$. The result is similar to that in the 3D case shown in Fig.~\ref{fig:suscep}(a); the susceptibility $\chi_{\rm 2D}^{zz}$ obeys the Curie law at high $T$ and exhibits the Van Vleck-type paramagnetism with $\chi_0' = \lim_{T \to 0} \chi_{\rm 2D}^{zz} = 1/2$ at low $T$, after showing a hump at $T\sim 1$. 
In this 2D case, however, $\chi_{\rm 2D}^{zz} - \chi_0' \propto \exp(-2\beta J_{\rm 2D})$ at low $T$, which is consistent with the lowest excitation gap $2J_{\rm 2D}$ by a single flip of $\tilde{B}_p$ allowed in the absence of the local constraints. 
Furthermore, there is no singularity in the susceptibility unlike the 3D case because of the absence of phase transition at finite $T$ in 2D.

\subsection{Dynamical spin correlation function}\label{sec:dynam-corr-funct}
\begin{figure}[t]
\begin{center}
\includegraphics[width=\columnwidth,clip]{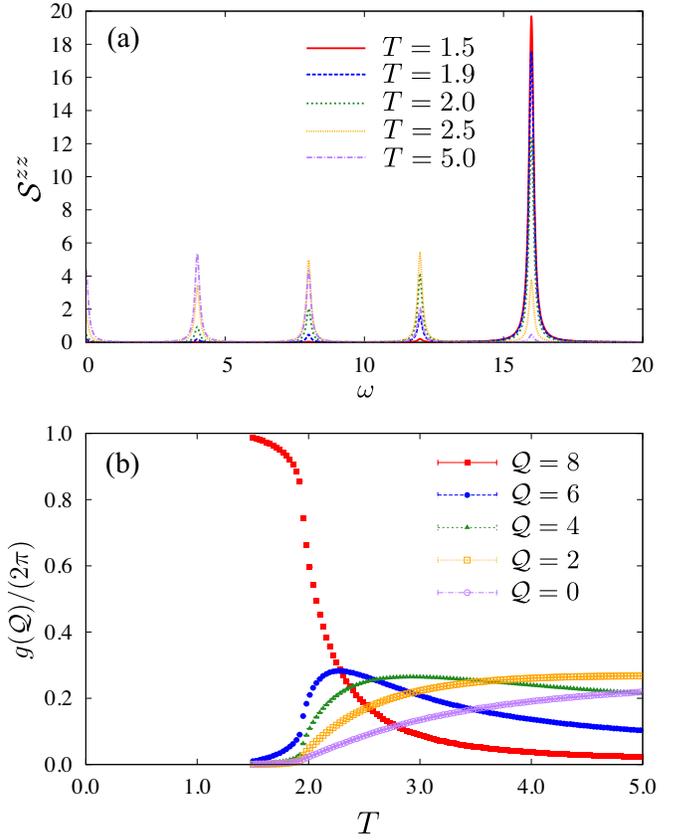}
\caption{(Color online) 
(a) The dynamical spin correlation function ${\cal S}^{zz}$ as a function of $\omega$ at several $T$. $\delta$-functions in Eq.~(\ref{eq:9}) are shown by Lorentzian functions with the width of $0.1$. 
(b) $T$ dependence of the intensities $g({\cal Q})$ of the peaks in ${\cal S}^{zz}$. The definition of $g({\cal Q})$ is given in the text. The system size is $L=30$.
}
\label{fig:dynamics}
\end{center}
\end{figure}

In this section, we calculate the dynamical spin correlation function which is observable in the inelastic neutron scattering and the resonant inelastic x-ray scattering measurements. 
The dynamical spin correlation function is defined as
\begin{align}
 {\cal S}_s^{zz}(\bm{q},\omega)=\frac{1}{N}\sum_{ij}\int_{-\infty}^{\infty}dt e^{i\omega t-i\bm{q}\cdot(\bm{r}_m-\bm{r}_n)}\means{S_i^z(t) S_j^z},
\end{align}
where $S_i^z(t)=e^{i{\cal H}t}S_i^z e^{-i{\cal H}t}$. In the limit of $J_z\gg J$, the dynamical spin correlation function can be described in terms of the PS $\bm{\tau}$ as
\begin{align}
 {\cal S}^{zz}(\bm{q},\omega)&=2{\cal S}_s^{zz}(\bm{q},\omega)\nonumber\\
&=\frac{1}{N_d}\sum_{mn}\int_{-\infty}^{\infty}dt e^{i\omega t-i\bm{q}\cdot(\bm{r}_m-\bm{r}_n)}\means{\tau_m^z(t) \tau_n^z},
\end{align}
where $\tau_m^z(t)=e^{i{\cal H}_{\rm eff}t}\tau_m^z e^{-i{\cal H}_{\rm eff}t}$. On the other hand, in a similar manner to Eq.~(\ref{eq:5}), the dynamical susceptibility is obtained in the Matsubara representation as
\begin{align}
 \chi^{zz}(i\omega_p)&=\frac{1}{N_d}\sum_{mn}
\int_0^\beta d\lambda \means{e^{\lambda{\cal H}_{\rm eff}}
\tau_m^z e^{-\lambda{\cal H}_{\rm eff}}
\tau_n^z
}e^{i\omega_p \lambda}\nonumber\\
&=\frac{1}{N_d}\sum_{m}\means{\tilde{f}({\cal Q}_m,i\omega_p)},
\end{align}
where $\omega_p=2\pi p T$ with an integer $p$ is the Matsubara frequency and
\begin{align}
 \tilde{f}({\cal Q}_m,i\omega_p)
=\int_0^\beta d\lambda e^{-2\lambda{\cal Q}_m+i\omega_p\lambda}=\frac{e^{-2{\cal Q}_m \beta}-1}{i\omega_p-2 {\cal Q}_m}.
\end{align}
Note that $\chi^{zz}(i\omega_p)$ does not depend on the momentum $\bm{q}$ because there is no spatial correlation in terms of $\tau^z$ in the effective Hamiltonian ${\cal H}_{\rm eff}$, as mentioned in Sec.~\ref{sec:magn-susc}. By using the fluctuation-dissipation theorem, the dynamical spin correlation function is evaluated by $\chi^{zz}$ as follows:
\begin{align}
 {\cal S}^{zz}(\bm{q},\omega)&=\frac{2}{1-e^{-\beta\omega}}{\rm Im} \chi^{zz}(\omega+i\eta)\label{eq:f-d_theorem_1}\\
&=\frac{1}{N_d}\sum_{m}\means{2\pi\delta(\omega-2{\cal Q}_m)}.\label{eq:9}
\end{align}
Thus, the dynamical spin correlation function is also momentum independent: ${\cal S}^{zz}(\bm{q},\omega) = {\cal S}^{zz}(\omega)$. Moreover, from Eq.~(\ref{eq:9}), ${\cal S}^{zz}(\omega)$ is given by $\delta$-functional peaks at $\omega=2{\cal Q}_{m}=0,4,8,12,16$.

Figure~\ref{fig:dynamics}(a) shows MC results for the dynamical spin correlation function as a function of $\omega$ at several $T$. The results show $\delta$-functional peaks as expected. At high $T$, intensities of the low energy peaks are stronger than those of high energy peaks. With deceasing $T$, the peak at the highest energy $\omega=16$ grows near $T_c$, whereas the intensities of the other peaks decrease. Figure~\ref{fig:dynamics}(b) shows $T$ dependence of the peak intensities in ${\cal S}^{zz}$. We define $g({\cal Q})$ so that the dynamical correlation function is written as 
\begin{align}
 {\cal S}^{zz}=\sum_{\cal Q}g({\cal Q})\delta(\omega-2{\cal Q}).\label{eq:11}
\end{align}
 As shown in Fig.~\ref{fig:dynamics}(b), the intensity of the highest energy peak, $g(8)$, monotonically increases with decreasing $T$, and rapidly grows near $T_c$ with showing the saturation to $2\pi$ at lower $T$. This result comes from the fact that, in the ground state, all $B_p$ are equal to $+1$ and all ${\cal Q}_m$ take 8. 

On the other hand, the other components show humps at a temperature above $T_c$ and decrease rapidly near $T_c$ with deceasing $T$. This nonmonotonic $T$ dependence originates from the intersite correlation of $B_p$ on the eight-site loops shown in Fig.~\ref{fig:eight_loop}.
This is in contrast to the 2D case. In the effective model given by Eq.~(\ref{eq:10}) for the 2D Kitaev model on the honeycomb lattice, the dynamical correlation function is given by
\begin{align}
 {\cal S}^{zz}_{\rm 2D}=\sum_{{\cal Q}=0,2}\tilde{g}({\cal Q})\delta(\omega-2{\cal Q}),
\end{align}
where $\tilde{g}(2)=2\pi(1+\tanh\beta J_{\rm 2D})^2/4$ and $\tilde{g}(0)=2\pi(1-\tanh^2\beta J_{\rm 2D})/2$. This result indicates that both $g(2)$ and $g(0)$ show monotonic $T$ dependence, while the former increases and the latter decreases as $T$ decreases. The $\delta$-functional contribution in the dynamical correlation function was recently obtained in Ref.~\onlinecite{Knolle2013}.

\subsection{Loop configuration}\label{sec:loop-configuration}

While a continuous phase transition is usually associated with symmetry breaking, the present model does not show any apparent symmetry breaking below $T_c$; it is hard to construct an order parameter in terms of the local variables $B_p$. Here, we try to characterize the phase transition by focusing on the global quantities, closed loops composed of the sites with $B_p=-1$. From the observation in MC samples that loops interpenetrating the system from one side of the surface to the opposite side are dominantly excited above $T_c$, we calculate the flux density defined by 
\begin{align}
\bar{\phi}^2/L=
\frac1L \sum_i \sum_{\mu=1}^3(\phi_i^\mu)^2.
\label{eq:fluxdensity}
\end{align}
Here, we define $\phi_i^\mu$ by the path integral $\phi_i^\mu=\oint_{{\cal C}_i}\bm{a}_\mu\cdot d\bm{s}/L$ for each loop ${\cal C}_i$ ($\bm{a}_\mu$ is a primitive translation vector of the pyrochlore lattice). A similar quantity was discussed in the 3D classical dimer model to characterize the high-$T$ Coulomb phase.~\cite{Alet2006} Since 
intersections of loops are allowed in the present model [see Fig.~\ref{fig:pyrochlore}(a)], we need to divide each intersection in order to identify the individual loop ${\cal C}_i$. Although there is an arbitrariness in the way to separate the loops as well as to assign the direction $d\bm{s}$, we confirm that $\langle \bar{\phi}^2 \rangle$ does not depend on the way.

\begin{figure}[t]
\begin{center}
\includegraphics[width=\columnwidth,clip]{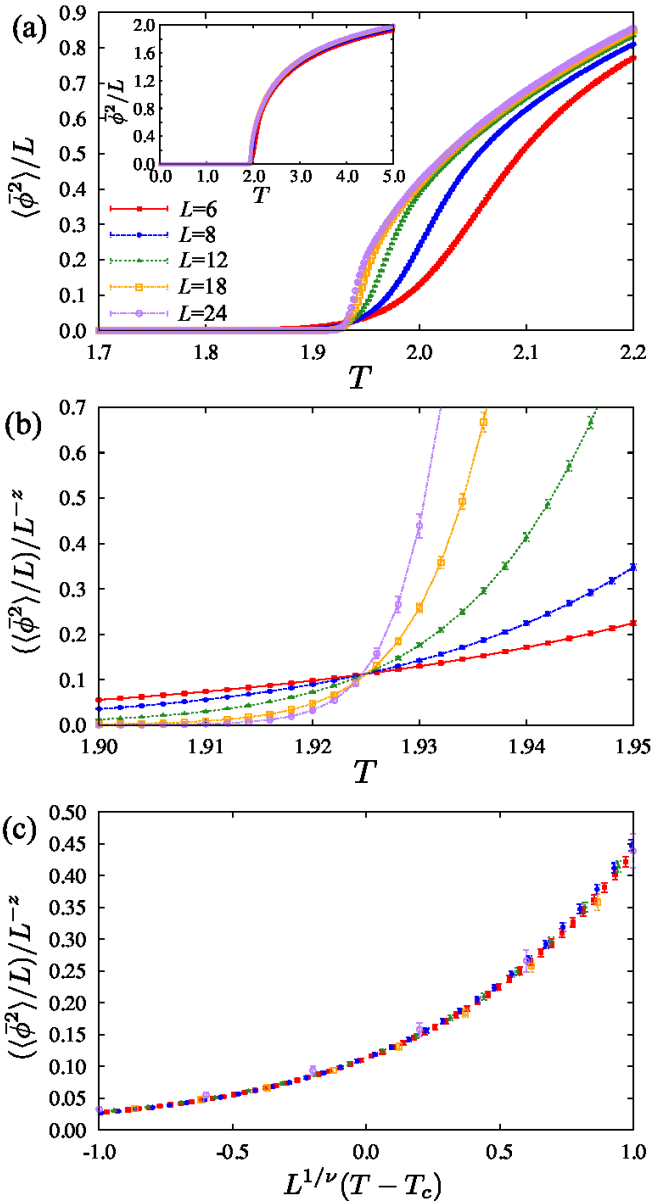}
\caption{(Color online) 
(a) MC results for $\means{\bar{\phi}^2}/L$. The inset in (a) shows $\means{\bar{\phi}^2}/L$ in a wide $T$ range. (b) $T$ dependence of $(\means{\bar{\phi}^2}/L)/L^{-z}$ and (c) scaling plot for $\means{\bar{\phi}^2}/L$. We assume $z=1$ in both (b) and (c). See the text for details.
}
\label{fig:scaling}
\end{center}
\end{figure}

\begin{figure}[t]
\begin{center}
\includegraphics[width=\columnwidth,clip]{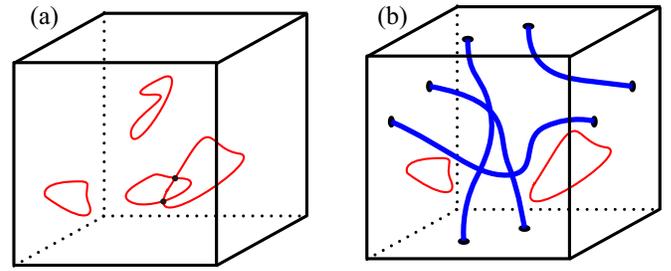}
\caption{(Color online) 
Schematic pictures of (a) the loop configuration below $T_c$ and (b) that above $T_c$. The thin red (thick blue) lines represent the loops with zero (nonzero) flux.}
\label{fig:loop}
\end{center}
\end{figure}

Figure~\ref{fig:scaling}(a) shows $T$ dependence of $\means{\bar{\phi}^2}/L$. This quantity exhibits an anomaly at $T_c$; while $\means{\bar{\phi}^2}/L$ is essentially zero below $T_c$, it becomes nonzero above $T_c$ in the thermodynamic limit. The behavior is similar to that in the classical dimer model,~\cite{Alet2006} which suggests that the high-$T$ paramagnetic phase has a Coulombic nature because of the constraints. The results indicate that the phase transition is characterized by the flux density. The flux $\phi_i^\mu$ corresponds to the winding number of the loop ${\cal C}_i$ along the vector $\bm{a}_\mu$. Hence, the result of $\means{\bar{\phi}^2}/L$ in Fig.~\ref{fig:scaling}(a) indicates that, below $T_c$, the winding numbers vanish and only short loops are excited, as schematically shown in Fig.~\ref{fig:loop}(a). On the other hand, the extended loops, which contribute to nonzero winding numbers, are generated above $T_c$ [see Fig.~\ref{fig:loop}(b)]. Therefore, the phase transition can be characterized by the flux in terms of the loop model; namely, the zero (non-zero) flux phase is associated with the low-$T$ QSL (high-$T$ paramagnetic) phase.

Despite the lack of an appropriate order parameter, we analyze the critical behavior of the phase transition from the paramagnetic side by using $\means{\bar{\phi}^2}$. Following the argument for the classical dimer model,~\cite{Alet2006} we assume the scaling $\means{\bar{\phi}^2}/L=L^{-z}f(L^{1/\nu}(T-T_c))$, where $z$ is a scale exponent, $\nu$ is a critical exponent for correlation length, and $f(T)$ is a scaling function. Figure~\ref{fig:scaling}(b) shows $T$ dependence of $(\means{\bar{\phi}^2}/L)/L^{-z}$ with $z=1$. The data for different $L$ cross with each other at the same point. This result indicates that the assumption $z=1$ is reasonable also for the present case. In addition, the scaling plot is shown in Fig.~\ref{fig:scaling}(c). Here, assuming $z=1$, we optimize $T_c$ and $\nu$ so that all the data collapse onto a single universal function. The analysis gives the estimates $T_c=1.925(1)$ and $\nu=0.60(5)$. The value of $T_c$ is fairly consistent with $T_c^{(L \to \infty)}$ in the inset of Fig.~\ref{fig:mc}(b). In addition, the value of the critical exponent $\nu$ is consistent with that of the 3D Ising universality class. The values of $T_c$ and $\nu$ will be discussed in the next section \ref{sec:corr-ising-model}. 

The loop picture presented in Fig.~\ref{fig:loop} will be effective for understanding the nature of phase transition even in the parameter region apart from the limit of $J/J_z \to 0$ considered here. A finite $J$ brings about complicated interactions in the effective Hamiltonian from higher-order contributions in terms of $J$, which lead to interactions between loops. Nevertheless, the higher-order terms do not alter the loop-like structure of excited states, since it is a direct consequence of the local constraint of $K_p$ in Eq.~(\ref{eq:8}) stemming from the basic spin algebra, which is imposed in the entire range of parameters. Given that the nature of transition is dominated by the global behavior of loop-like excitations in the limit of $J/J_z \to 0$ as sketched in Fig.~\ref{fig:loop}, it is reasonable to expect that a similar transition takes place even at finite $J$. Such extension will be discussed elsewhere.

\subsection{Effective model without global constraints}\label{sec:corr-ising-model}

\begin{figure}[t]
\begin{center}
\includegraphics[width=\columnwidth,clip]{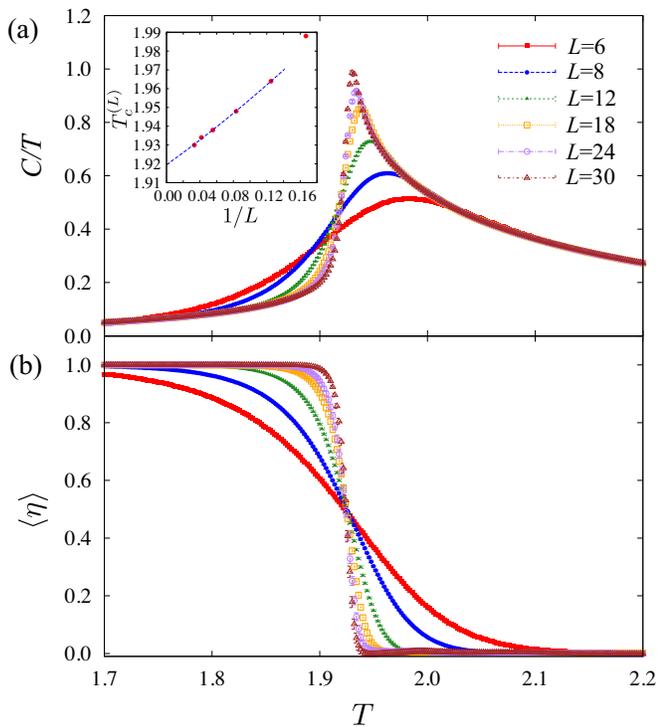}
\caption{(Color online) 
MC results for (a) $C/T$ and (b) the loop parity parameter $\mean{\eta}$ for the effective model in Eq.~(\ref{eq:2}) without the two global constraints.
The inset in (a) shows the peak temperature of $C$ as a function of $1/L$. The dotted curve represents the quadratic fit for 5 largest $L$. 
}
\label{fig:mc_loop}
\end{center}
\end{figure}

As described in Sec.~\ref{sec:model}, the effective model in Eq.~(\ref{eq:2}) is subject to both local and global constraints for the variables $B_p$. The local constraint restricts the excited states in the form of closed loops of $B_p=-1$ sites. It plays a decisive role on the finite-$T$ phase transition, as demonstrated in the previous sections. On the other hand, the role of the global constraints (i) and (ii) is not obvious. In this section, we examine it by studying the model with omitting the global constraints.

Figure~\ref{fig:mc_loop}(a) shows $T$ dependence of the specific heat divided by $T$ for the model in the absence of the global loops. In the MC simulation, we allow MC update by flipping a single six-site loop as well as the global update along an extended loop. Similar to the result in Fig.~\ref{fig:mc}(b), the specific heat exhibits divergent behavior, although the peak values are slightly smaller than those in Fig.~\ref{fig:mc}(b). The extrapolation of the peak temperature of $C$ to $L\to \infty$ gives an estimate of $T_c = 1.920(2)$, as shown in the inset of Fig.~\ref{fig:mc_loop}(a).
The estimate coincides well with that obtained in the inset of Fig.~\ref{fig:mc}(b). Furthermore, we confirm that the scaling of the flux density $\langle \bar{\phi}^2\rangle/L$ also gives the same critical exponents, as in Sec.~\ref{sec:loop-configuration} (not shown). These results indicate that the global constraints (i) and (ii) are not relevant to the thermodynamics in the present effective model; the finite-$T$ phase transition takes place at the same $T_c$ with the same critical properties.

For further analysis from the viewpoint of the loop topology, let us introduce the parity variable $\eta=\eta_x\eta_y\eta_z$ for loop configurations. Here we define $\eta_l=(-1)^{n_l}$, 
where $n_l$ ($l=x,y,z$) is the number of extended loops in the $l$ direction. 
The parity $\eta$ distinguishes topologically different sectors of the loop configurations in the current model where loops are allowed to intersect with each other. In addition, $\eta$ is written by the flux of loops as $\eta=e^{i\pi\bar{\phi}^2}$.
Note that $\eta$ is always +1 in the presence of the global constraints.

Figure~\ref{fig:mc_loop}(b) shows the numerical result for $T$ dependence of $\mean{\eta}$. At high $T$, loop configurations with odd and even parity appear in the same weight, and hence, $\mean{\eta}$ vanishes. On the other hand, below $T_c$, there are only short loops in the system as presented in Sec.~\ref{sec:loop-configuration}, and hence, $\mean{\eta}$ takes $+1$. In the vicinity of $T_c$, $\mean{\eta}$ changes rapidly and the slope at $T_c$ increases with increasing the system size. The result implies that $\mean{\eta}$ changes discontinuously from 0 to 1 at $T_c$ in the thermodynamic limit. Thus, this quantity appears to work as an order parameter for this phase transition, although the discontinuous jump looks incompatible with the continuous transition. As shown in Appendix~\ref{sec:2d-loop-model}, however, the similar parity variable in a loop model on the 2D square lattice is indeed analytically shown to be a step function $\theta(\tilde{T}_c-T)$ in the thermodynamic limit, where $\tilde{T}_c$ is the critical temperature in the 2D loop model, though the transition is continuous in the Ising universality class.

Finally, we discuss a correspondence between our model and the Ising model.~\cite{Kamiya_pc} If the global constraints (i) and (ii) are omitted, the effective loop model given by Eq.~(\ref{eq:2}) becomes equivalent to the Ising model with NN interactions on the diamond lattice; by identifying the contributions from loops of length $\ell$ with the $\ell$-th order terms in the high-$T$ series expansion of the Ising model on the diamond lattice, the partition function of the loop model, $Z(\beta)$, is associated with that of the Ising model on a diamond lattice, $Z_{\rm Ising}(\beta)$, as
\begin{align}
 Z_{\rm Ising}(\beta)=2^{N_d}(e^{\beta}\cosh\beta)^{zN_d/2} Z\Big{(}-\frac12 \ln \tanh\beta\Big{)},\label{eq:3}
\end{align}
where $z=4$ is the coordination number of the diamond lattice.
 From this relation, we obtain the critical temperature for the present loop model as
\begin{align}
T_c = -\Bigl[\frac{1}{2}\ln\tanh
\Big{(}\frac{1}{T_c'}\Big{)} \Bigr]^{-1},
 \label{eq:7}
\end{align}
where $T_c'$ is the critical temperature of the Ising model on the diamond lattice. Using this relation and the previous result $T_c' = 2.7042(2)$,~\cite{Essam1963,Gaunt1973} we obtain $T_c = 1.9249(2)$. This value is consistent with our results for the effective model with and without global constraints. In addition, as discussed above, the critical exponent $\nu$ obtained from the scaling analysis of the flux density is also consistent with that of the 3D Ising universality class. These results confirm the validity of our characterization of the phase transition, although the correspondence to the Ising model is ensured only in the absence of the global constraints.
The correspondence between the $Z_2$ spin liquid in a 3D toric code model and the 3D Ising model was also discussed in Ref.~\onlinecite{Castelnovo2008}.

We note, however, that this correspondence does not describe the critical exponent of magnetic susceptibility $\chi^{zz}$ correctly. This is reasonable because $\chi^{zz}$ of the current 3D Kitaev model is defined through the correlation function of PS operators $\tau_m^z$, which does not show long-range ordering below $T_c$, in contrast to Ising spins in the 3D Ising model. To obtain the correct critical behavior of $\chi^{zz}$, we need the explicit calculations in terms of $\tau_m^z$, as discussed in Sec.~\ref{sec:magn-susc}.


\section{Concluding remarks}\label{sec:concluding-remarks}

To summarize, we have investigated thermodynamic properties of the 3D Kitaev model on the hyperhoneycomb lattice. Focusing on the anisotropic case, in which the ground state is exactly shown to be a gapped QSL, we have studied the finite-$T$ properties in the anisotropic limit by MC simulation. We have found a phase transition between QSL and paramagnetic phase at a finite $T$. 
We have characterized the phase transition by the flux density of loop excitations, which is nonzero in the high-$T$ paramagnet and zero in the low-$T$ QSL. As a result, the transition turns out to be second order and belongs to the 3D Ising universality class. 
The local constraints in the Ising-type variables play an essential role in the occurrence of the phase transition at a finite $T$.

We have also obtained the static magnetic susceptibility and dynamical spin correlation function. The magnetic susceptibility obeys the Curie law at high $T$, while it shows Van Vleck-type constant behavior at low $T$ after showing a hump above the critical temperature $T_c$. At $T_c$, the magnetic susceptibility is continuous, but its $T$ derivative shows divergence. 
The dynamical spin correlation is momentum independent but shows quantized peaks as a function of excitation energy.
The peak at the highest energy rapidly increases near $T_c$ as decreasing $T$, while some of other components exhibit humps before decreasing near $T_c$. These findings provide a new insight into the thermodynamics of QSLs and will stimulate further studies of the new state of matter. 

Finally, we make some remarks including the relation to real materials. In our finding of the stable QSL at finite $T$, the constraints that restrict the excitations in the form of closed loops play an essential role. While our study has been limited for the gapped QSL in the limit of strong anisotropy, a finite-$T$ transition might also be seen widely, even in the gapless QSL region, as similar constraints are imposed in the entire parameter region in the current 3D Kitaev model. Such extension for generic parameters including the isotropic case with $J_x=J_y=J_z$ will be reported elsewhere. Meanwhile, the model possibly provides the simplest reference to a recently-found hyperhoneycomb compound $\beta$-Li$_2$IrO$_3$.~\cite{Takagi_pc} Our numerically-exact results will give a firm ground for understanding of the thermodynamic properties of the compound.

{\it Note added}:
In the completion of our study, we received two preprints on a closely related topic, one of which was published during the editorial process.~\cite{Lee2014,Kimchi1309}

\begin{acknowledgments}
The authors thank Y.~Kamiya and K.~Hukushima for helpful discussions. They are also grateful to H.~Takagi for the information on the experiments. One of the authors (J.N.) acknowledges the financial support from the Japan Society for the Promotion of Science. This research was supported by KAKENHI (No. 24340076 and No. 24740221), the Strategic Programs for Innovative Research (SPIRE), MEXT, and the Computational Materials Science Initiative (CMSI), Japan. Parts of the numerical calculations are performed in the supercomputing systems in ISSP, the University of Tokyo.
\end{acknowledgments}

\appendix

\section{Derivation of Van Vleck-type paramagnetism}\label{sec:derivation-van-vleck}

In this appendix, we evaluate the value of the magnetic susceptibility in Eq.~(\ref{eq:chi^zz_Q}) in the limit of $T\rightarrow 0$. The susceptibility is rewritten into
\begin{align}
 \chi^{zz}=\sum_{\cal Q}\means{P({\cal Q})}f({\cal Q}),\label{eq:12}
\end{align}
where $P({\cal Q})=\sum_m p_m({\cal Q})/N_d$ 
and $p_m({\cal Q})$ is the probability such that ${\cal Q}_m$ is equal to ${\cal Q}$; $f({\cal Q})$ is given in Eq.~(\ref{eq:fQ}). By definition, $\sum_{\cal Q}P({\cal Q})=1$ is satisfied. 
In the low-$T$ limit, the canonical average $\means{\cal O}$ can be expanded as
\begin{align}
 \means{{\cal O}}=\means{{\cal O}}_0+D_1 (\means{{\cal O}}_1-\means{{\cal O}}_0)e^{-\beta \Delta E_1}+\cdots,
\end{align}
where $\means{{\cal O}}_0$ and $\means{{\cal O}}_1$ are the microcanonical averages in the ground state and the first excited state, respectively. Here, we assume that the ground state is not degenerate. $D_1$ is the number of states  in the first excited state and $\Delta E_1$ is the energy difference between the ground state and the first excited state. In the present case, $D_1=6L^3$ and $\Delta E_1=16$ because the first excited state is given by an eight-site loop formed by $B_p=-1$ sites, as discussed in Sec.~\ref{sec:magn-susc}. 
Then, the susceptibility in the low-$T$ limit is written as
\begin{align}
 \chi^{zz}&=\sum_{\cal Q}f({\cal Q})[\means{P({\cal Q})}_0\nonumber\\
&\ \ +6L^3\{\means{P({\cal Q})}_1-\means{P({\cal Q})}_0\}e^{-16\beta} +\cdots].
\end{align}
Since the ground state is the state where all $B_p$ are $+1$, the microcanonical average $\means{P({\cal Q})}_0$ is $\delta_{{\cal Q},8}$. In addition, for negative ${\cal Q}$, $f({\cal Q})$ diverges as $\frac{-1}{2{\cal Q}}e^{-2\beta{\cal Q}}$ in the low-$T$ limit. Because $f({\cal Q}=-8)$ exhibits the strongest divergence, the value of $\chi^{zz}$ in the limit of $T\rightarrow 0$ is given by
\begin{align}
 \chi_0&= \lim_{T\to 0} \chi^{zz}(T)\nonumber\\ 
 &=f({\cal Q}=8)+6L^3\means{P({\cal Q}=-8)}_1 f({\cal Q}=-8)e^{-16\beta}.
\end{align}
As the divergence of $f({\cal Q}=-8)$ cancels with $e^{-16\beta}$, we obtain
\begin{align}
\chi_0=\frac{1}{16}+\frac{1}{16}6L^3 \means{P({\cal Q}=-8)}_1.
\label{eq:chi0_final}
\end{align}

The value of $\means{P({\cal Q}=-8)}_1$ is evaluated as follows.
There are six kinds of eight-site loops in the pyrochlore lattice. Among them, the two kinds of loops shown in Fig.~\ref{fig:eight_loop} contribute to the susceptibility. A set of these two kinds of loops is written as ${\cal L}_{\chi}$. 
Since the present system has a translational symmetry, $\means{P({\cal Q}=-8)}_1$ is equivalent to $\means{p_m({\cal Q}=-8)}_1$ for a certain $m$. The probability of finding ${\cal L}_{\chi}$ in all eight-site loops is $(2L^3)/(6L^3)=1/3$, and the eight-site loop given by ${\cal Q}_m=-8$ is one of the eight-site loops in ${\cal L}_{\chi}$, the number of which is $(2L^3)$. Thus, $\means{p_m({\cal Q}=-8)}_1$ is evaluated as $(1/3)/(2L^3)$.
Then, the Van Vleck-type component of the susceptibility in Eq.~(\ref{eq:chi0_final}) is evaluated as $\chi_0=1/8$.
This asymptotic value is indeed observed in our MC simulation, as shown in Fig.~\ref{fig:suscep}(a).

There is another way to calculate the contribution at ${\cal Q}=-8$ in Eq.~(\ref{eq:12}). From Eq.~(\ref{eq:f-d_theorem_1}), the imaginary part of the dynamical susceptibility is written as
\begin{align}
 \frac{1}{\pi}\frac{{\rm Im}\chi^{zz}(\omega)}{\omega}=\sum_{\cal Q}f(Q)\means{P({\cal Q})}\delta(\omega-2{\cal Q}),
\end{align}
where we use the relation $g({\cal Q})/(2\pi)=\means{P({\cal Q})}$ which is obtained from the Kramers-Kronig relation for $\chi^{zz}$ given in Eq.~(\ref{eq:12}).
Since the imaginary part of the susceptibility is an odd function for $\omega$, $f({\cal Q}=-8)\means{P({\cal Q}=-8)}=f({\cal Q}=8)\means{P({\cal Q}=8)}=1/16$ is satisfied in the limit of $T\rightarrow 0$.

\section{2D loop model}\label{sec:2d-loop-model}

\begin{figure}[t]
\begin{center}
\includegraphics[width=1\columnwidth,clip]{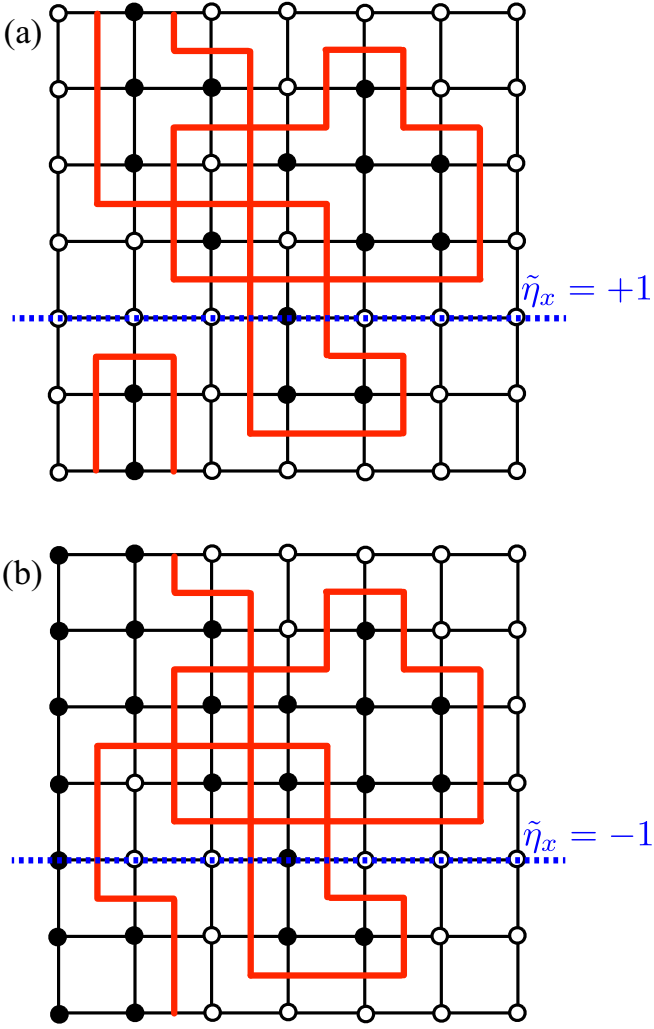}
\caption{(Color online) 
Spin configurations of the Ising model on the square lattice with (a) periodic-periodic and (b) antiperiodic-periodic boundary conditions. The system size is $6 \times 6$, and the leftmost (bottom) sites are equivalent to the rightmost (top) sites. Filled and open circles represent up- and down-spin states, respectively. Loop configurations corresponding to each spin configuration are represented by thick red lines. The parity $\tilde{\eta}_x$ takes different values corresponding to different boundary conditions in the $x$ direction.
}
\label{fig:ising}
\end{center}
\end{figure}

In this appendix, we derive $T$ dependence of the loop parity in a loop model defined on the 2D square lattice. In this model, the energy is given by the sum of the loop lengths and loops can intersect with each other in the same manner of the 3D loop model discussed in the main text. Since loops are regarded as domain walls in the Ising model as schematically shown in Fig.~\ref{fig:ising}, the partition function of the 2D loop model is calculated from that of the Ising model on the square lattice. For the 2D loop model, we introduce the parity variable $\tilde{\eta}=\tilde{\eta}_x\tilde{\eta}_y$, where $\tilde{\eta}_x$ $(\tilde{\eta}_y)$  is the parity of the number of loops which intersect with the $x$ $(y)$ axis: $\tilde{\eta}$ is the parity of the number of extended loops. 
In terms of the Ising model, $\tilde{\eta}_l$ represents the parity defined by the number of domain walls along the $l$ direction. This parity is determined by the boundary condition; the even parity $\tilde{\eta}_l=+1$ is realized in the periodic boundary condition and the odd parity $\tilde{\eta}_l=-1$ is realized in the antiperiodic boundary condition (see Fig.~\ref{fig:ising}). Therefore, the thermal average of $\tilde{\eta}$ is given by
\begin{align}
 \mean{\tilde{\eta}}=\frac{Z_{\rm pp}+Z_{\rm aa}-Z_{\rm pa}-Z_{\rm ap}}{Z_{\rm pp}+Z_{\rm aa}+Z_{\rm pa}+Z_{\rm ap}},
\end{align}
where $Z_{\rm pp}$, $Z_{\rm aa}$, $Z_{\rm pa}$, and $Z_{\rm ap}$ are the partition functions of the Ising model on the $L\times L$ square lattice with the periodic-periodic, antiperiodic-antiperiodic, periodic-antiperiodic, and antiperiodic-periodic boundary conditions, respectively. These partition functions are given by~\cite{Kastening2002,Wu2002}
\begin{align}
 Z_{\rm pp}&=\frac{1}{2}[C_o+S_o+C_e-S_e'{\rm sgn}(1-\sinh\beta)],\\
 Z_{\rm aa}&=\frac{1}{2}[-C_o+S_o+C_e+S_e'{\rm sgn}(1-\sinh\beta)],\\
 Z_{\rm pa}&=\frac{1}{2}[C_o-S_o+C_e+S_e'{\rm sgn}(1-\sinh\beta)],\\
 Z_{\rm ap}&=\frac{1}{2}[C_o+S_o-C_e+S_e'{\rm sgn}(1-\sinh\beta)],
\end{align}
where
\begin{align}
 \ln C_o&=\frac{1}{2}\sum_{p=0}^{L-1}\sum_{q=0}^{L-1}\ln\Biggl[\cosh^2\beta
\nonumber\\
&\ \ \ \ 
-\sinh\beta \left(
\cos\frac{(2p+1)\pi}{L}
+\cos\frac{(2q+1)\pi}{L}
\right)\Biggr],\\
\ln S_o&=\frac{1}{2}\sum_{p=0}^{L-1}\sum_{q=0}^{L-1}\ln\Biggl[\cosh^2\beta
\nonumber\\
&\ \ \ \ 
-\sinh\beta \left(
\cos\frac{2p\pi}{L}
+\cos\frac{(2q+1)\pi}{L}
\right)\Biggr],\\
\ln C_e&=\frac{1}{2}\sum_{p=0}^{L-1}\sum_{q=0}^{L-1}\ln\Biggl[\cosh^2\beta
\nonumber\\
&\ \ \ \ 
-\sinh\beta \left(
\cos\frac{(2p+1)\pi}{L}
+\cos\frac{2q\pi}{L}
\right)\Biggr],\\
\ln S'_e&=\frac{1}{2}\sum_{p=0}^{L-1}\sum_{q=0}^{L-1}\ln\Biggl[\cosh^2\beta
\nonumber\\
&\ \ \ \ 
-\sinh\beta \left(
\cos\frac{2p\pi}{L}
+\cos\frac{2q\pi}{L}
\right)\Biggr].
\end{align}
Here, the exchange constant of Ising model is chosen to be unity. In the thermodynamic limit, since $C_o=S_o=C_e=S_e'$ is satisfied, $\mean{\tilde{\eta}}$ is calculated as
\begin{align}
\mean{\tilde{\eta}}&=\frac{S_o+C_e-C_o-S'_e {\rm sgn}(1-\sinh\beta)}{S_o+C_e+C_o+S'_e {\rm sgn}(1-\sinh\beta)}\nonumber\\
&=\frac{1-{\rm sgn}(T-\tilde{T}_c)}{3+{\rm sgn}(T-\tilde{T}_c)}=\theta(\tilde{T}_c-T),\label{eq:6}
\end{align}
where $\theta$ is a step function and $\tilde{T}_c$ is the critical temperature of the 2D Ising model determined by $\sinh(1/\tilde{T}_c)=1$.~\cite{Kastening2002,Wu2002}
 This result indicates that $\tilde{\eta}$ is $1$ ($0$) for $T<\tilde{T}_c$ ($T>\tilde{T}_c$), and exhibits a discontinuous change at $\tilde{T}_c$. 
Although the phase transition in this system is a continuous transition, the parameter $\means{\tilde{\eta}}$ is discontinuous at $T_c$ and does not show the $T$ dependence except for $T_c$. This behavior originates from the fact that $\tilde{\eta}$ cannot be written by a local quantity but describes the topology of loops in the system. Our result indicates that a topological aspect of the second-order phase transition is characterized by a discontinuous variable $\tilde{\eta}$.



\end{document}